\documentclass[preprint2]{aastex}

\usepackage{url}
\usepackage{amsmath}
\renewcommand{\=}{\linebreak[0]=\linebreak[0]}

\newcommand{\lb}[1]{\linebreak[0]\texttt{#1}\linebreak[0]}

\begin{document}

\title{IEAD: A Novel One-Line Interface to Query Astronomical Science
  Archives} 

\author{Marco Lombardi}
\affil{University of Milan, Department of Physics, via Celoria 16,
  I-20133 Milan, Italy}

\begin{abstract}
  In this article I present IEAD, a new interface for astronomical
  science databases.  It is based on a powerful, yet simple, syntax
  designed to \textit{completely\/} abstract the user from the
  structure of the underlying database.  The programming language
  chosen for its implementation, JavaScript, makes it possible to
  interact directly with the user and to provide real-time
  information on the parsing process, error messages, and the name
  resolution of targets; additionally, the same parsing engine is used
  for context-sensitive autocompletion.  Ultimately, this product
  should significantly simplify the use of astronomical archives,
  inspire more advanced uses of them, and allow the user to focus on
  \textit{what\/} scientific research to perform, instead of on
  \textit{how\/} to instruct the computer to do it.
\end{abstract}


\section{Introduction}

Nowadays, a well-maintained and easily accessible data archive is
critical to the success of a mid-to-large telescope facility.  This is
best appreciated if one looks at the large amount of \textit{pure\/}
archival articles, i.e\ articles written using data from observations
that were not proposed by any of the authors.  For example, as noted
already by \citet{2006STECF..40....6W}, approximately half of the
publications based on Hubble Space Telescope data are purely archival.
Furthermore, archival research is of course largely dominant for
surveys and for dedicated telescopes, such as the Sloan Digital Sky
Survey \citep{2000AJ....120.1579Y} or the Two Micron All Sky Survey
\citep{2006AJ....131.1163S}.

Astronomical archives are very complex.  On one hand, the internal
structure of the database is often complicated by the need to collect
intrinsically different kinds of data, taken for different purposes by
different instruments.  On the other hand, the archive interface must
serve technical customers, the astronomers, who sometimes need to
perform very specific queries.  As a result, typical archive
interfaces (almost always accessible through a dedicated World Wide
Web page) are often plagued by a large list of different fields and
buttons to be able to accommodate queries from the most demanding (and
technically inclined) user.

Unfortunately, this also means that many archive interfaces are also
clumsy and unfriendly for the large majority of users.  For example, a
recent survey among the ESO Science Archive users
(\citealp{2006Msngr.125...41D}; see also the complete survey at
\url{http://archive.eso.org/archive/stats/survey/survey_results.html})
has shown that the most requested improvements for the archive
interface are the possibility to perform more complex queries (23\%),
an easier-to-use interface (20\%), and a less dense main-query page
(17\%).  Clearly, these three suggestions cannot be followed at the
same time using a classical interface.

As of recently, an increasing number of astronomical archive
interfaces now accept queries written in SQL or extensions of this
language (such as ADQL, \citealp{ADQL}).  Unfortunately, while
presenting a clean query page, these solutions require the user to
know the internal structure of the database used at a relatively deep
level and they ultimately make the database inaccessible to the less
technically-inclined users.  Additionally, since different
astronomical databases often have completely different structures, one
is forced to learn for each archive used very specific details that
are of no use in different contexts.

However, perhaps the most serious limitation of the currently
available interfaces to astronomical databases is the fact that they
force the user to express the query in a form that (in all cases)
strictly reflects the structure of the database.  Ideally, instead, the
user should focus on the science, and the interface should be flexible
enough to allow a direct formulation of the user wishes.

In this article I present IEAD, or Interface for the Exploration of
Astronomical Databases, a new query interface for astronomical science
archives that solves the limitations discussed above.  The concepts
presented here have been developed for the Hubble Space Telescope
(HST) science archive, but could be applied equally well without
significant modifications to any astronomical data archive (and the
software has been developed with this aim in mind).  IEAD is currently
available (integrated within a more standard query interface) at the
ST-ECF HST Archive (\url{http://archive.eso.org/archive/hst/search})
and at the CADC HST Archive
(\url{http://www.cadc.hia.nrc.gc.ca/hst/new}) under the name
``one-line query''.

The paper is organized as follows.  In Sect.~2 I present the basic
ideas behind the IEAD system and I briefly introduce its main
features.  A few user aids (including interactive parsing and
autocompletion) are discussed in Sect.~3.  The full general syntax is
described in detail in Sect.~4, and its implementation in Sect.~5.  In
Sect.~6 possible extensions of this research are presented together
with some of the difficulties that one might encounter.  Finally, the
paper is closed with a quick summary (Sect.~7).

\section{IEAD: the concept}
\label{sec:one-line-queries}

With the advent of powerful and ``smart'' search engines, we all are
used to the idea that simple interfaces should be provided to perform
recurrent tasks.  However, behind the apparent simplicity of these
search engines, there are a lot of complex tasks that are performed
behind the scenes: for example, some web search engines allow queries
to be formulated using natural languages, or with boolean operators,
or using special keywords to restrict the outputs.  In general, it
appears that the current focus in the development of search engines is
to adapt them the user needs, rather than to force the user to adapt
to their designs and limitations.

A few search engines are designed to be able to perform both simple
and (very) complex queries.  Examples can be found in many Google
products (the standard web search engine, but also the search
interface for messages in GMail or for RSS in the Google RSS reader)
and on many e-commerce sites (such as Amazon). In the astronomical
field, a similar but much simpler product can be found in the one-line
NASA ADS interface.\footnote{See \url{http://adswww.harvard.edu}.}\@
In a different context, the \texttt{get} script command of Aladin can
associate automatically a set of keywords (that can be entered in an
arbitrary order) with the server query vocabulary (see
\url{http://aladin.u-strasbg.fr/java/FAQ.htx#ToC99}).  It is along
these lines that I designed the IEAD search engine for the Hubble
Space Telescope archive.

Ideally, the ``perfect'' interface should be very intuitive to use and
little or no explanations should be needed; still, it should be
powerful enough to allow complex queries.  The interface should use a
simple syntax, or should accept queries formulated in a
\textit{natural language\/}.  Finally, the user should be able to
profit from the archive without knowing in detail the structure of the
database.

Natural language user interfaces are generally difficult to implement
and represent a very active field of research in computer science
\citep[e.g.][]{ART95, PEK03}.  In particular, a critical task in a
natural language interface is the \textit{entity identification\/},
i.e.\ the classification of the various terms present in a query.
Fortunately, astronomical data archives are very favorable in this
respect because many query terms can be uniquely associated with a
particular data field, a fact that makes it possible to have
unambiguous and still very simple queries.  For example, quantities
such as instrument names, camera names, filters, optical element types
(such as ``filter'' or ``grism'' or ``prism'') can assume only a fixed
set of single-word values.  Slightly more complex values, such as
principal investigator (PI) names or data type names (such as
``imaging'' or ``2D spectroscopy'') still assume values from a limited
set of single or multi-word values.  Real values, such as astronomical
coordinates, search radii, exposure times, exposure dates, or
observing wavelengths, can usually be easily disentangled from one
another by their format (say \textit{hh\/}:\textit{mm\/}:\textit{ss\/}
for Right Ascension vs. $\pm$\textit{dd\/}:\textit{mm\/}:\textit{ss\/}
for Declination) or from their units (\texttt{30arcsec} is probably a
search radius, while \texttt{2h} is likely an exposure time).
Finally, everything else, i.e.\ everything that are not recognized as
one of the fields mentioned above, has to be a target name, and this
assumption can then be verified using the SIMBAD
\citep{2000A&AS..143....9W} or NED \citep{1991ASSL..171...89H} name
resolvers.

In this simple approach, the string (queries are case insensitive)
$$
\texttt{acs m42 f775w}
$$
is immediately understood to be a query for all
\texttt{instrument\=acs} data taken with the \texttt{filter\=f775w}
around the (SIMBAD resolved) coordinates of \texttt{m42}.  This simple
example immediately highlights one of the main advantages of the IEAD
system over the other query interfaces commonly found for astronomical
databases: a level of abstraction is removed, and the user is free to
express the query in much more natural way.  This unique feature makes
the use of the database more direct, and ultimately makes database
research much easier by bringing the query language closer to the
astronomer.  Note that although the use of automatic entity
identification is not entirely new in the astronomical context (cf.\
the aforementioned NASA ADS query interface and the Aladin
\texttt{get} script command), IAED pushes this concept much forward:
now entire database queries, possibly composed of more than twenty
different kinds of entities, are automatically parsed.  Compared to
standard archive interfaces, as an additional bonus the user does not
need to look for the right quantities and enter different values in
the correct fields, but rather can mix together all values and still
obtain sensible results.

As a second example, the string
$$
\texttt{stis OII imaging planetary nebula}
$$
is interpreted as a query for all \texttt{instrument\=stis}
\texttt{data\_type\=imaging} data taken with the \texttt{filter\=OII}
for \texttt{description\="planetary nebula"}.  Finally, the example
$$
\texttt{nic3 2d spectroscopy hdf-n thompson >30min }
$$
can be used to select all \texttt{camera\=nic3} \texttt{data\_type="2d
  spectroscopy"} observations made by \texttt{pi=thompson} around the
\texttt{target=HDF-N}, with \texttt{exptime\lb{>}30min}.

\section{User aids}
\label{sec:user-aids}

Since the IEAD query interface is based on JavaScript, it runs
entirely on the user computer; additionally, the parser is custom
written, and is therefore fast enough to parse complex queries in real
time.  This makes it possible to complement the interface with two
user aids: an automatic display of the parsing result, and a smart the
autocompletion on the query.

The automatic display of the parsing result shows not only how every
word entered is interpreted, but also the way that the various
constraints are linked together.  For example, for the first example
discussed above the interface it shows
\begin{align*}
  & \texttt{(inst=acs and (target=m42[ok,HII] and} \\
  & \texttt{ box=600arcsec) and filter=f775w)}
\end{align*}
This text provides a wealth of information.  First, it is obvious that
\texttt{acs} is taken to be an instrument, \texttt{f775w} a filter,
and \texttt{m42} a target name.  Additionally, the target has been
correctly resolved and is a HII region (\texttt{[ok,HII]}); note that
the resolver (SIMBAD) and the meaning of the HII code (HII ionized
region) is visible if the user leaves the cursor over the HII text.
All terms are combined using the \texttt{and} boolean operator, and
the coordinate search is performed around the target name with the
default searching box, $10 \mbox{ arcmin}$.

In case of a parsing error the system is also able to show
immediately a descriptive error message and to indicate where the
error took place.  Similarly, the system also informs the user in real
time when a target cannot be resolved: this is a no-stopping error (in
the sense that the user can still type text in the field) since the
problem might rely on the target name to be incomplete; however, the
query will not be executed if the target cannot be resolved.

Additionally, when the user starts filling in the query field, the
system shows all possible completions.  The autocompletion is
automatically shown as soon as the last query text has at the end an
incomplete word (i.e., no completion is shown if the query text is
empty or if its last character is a space).  The autocompletion uses
the complete parsing engine, and therefore is fully context sensitive:
at any given moment the system knows exactly what are the possible
completion words, and can show them to the user.

Finally, it is worth mentioning that both the ST-ECF and the CADC HST
archive pages embed the IEAD system inside a traditional query form,
composed of different entries for each field.  In these pages each
entry from the traditional query form can be dragged and dropped into
the IEAD one-line interface.  This should help users to get acquainted
with the new interface and to enter there constraints for more hardly
to remember, uncommon keywords.

\section{General syntax}
\label{sec:general-syntax}

\begin{figure*}
\includegraphics[width=\hsize]{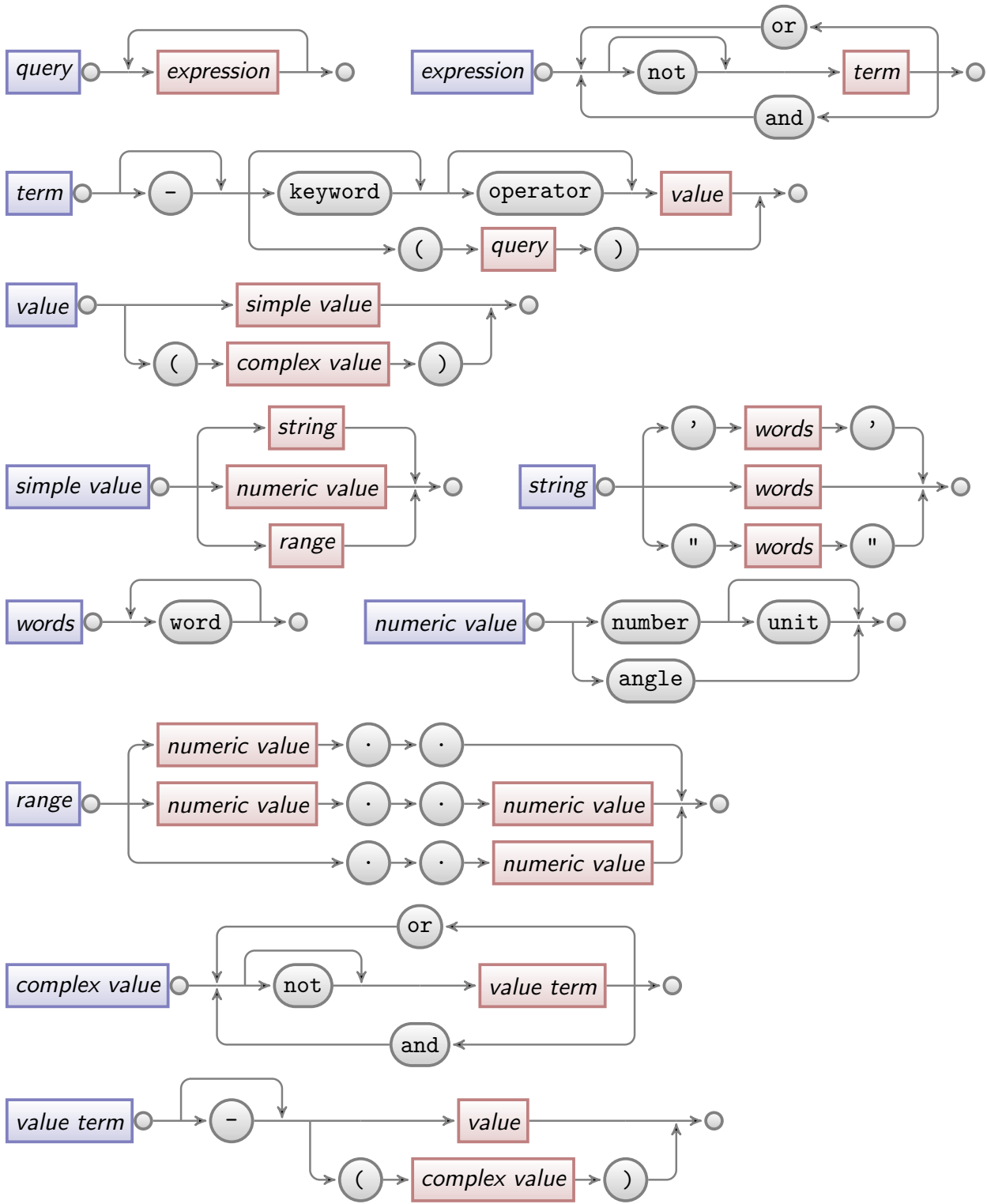}
\caption{Simplified syntax diagram for IEAD.  Non-terminal nodes are
  indicated using squared boxes, while terminal ones have rounded
  corners.\label{fig:1}}
\end{figure*}

The general syntax accepted by IEAD is represented in Fig.~\ref{fig:1}
and will be explained in detail in this section.  It is based on
\textit{qualified terms\/}, i.e.\ on a combination
\textit{keyword-operator-value\/} such as
$$
\texttt{instrument\=ACS}
$$
Both the keyword and the operator can be omitted: if the keyword is
omitted, it is inferred from the value (in the example above,
\texttt{ACS} is obviously an instrument); if the operator is omitted,
the ``equal'' (\texttt{=}) is assumed, unless the value is preceded
with a minus sign, in which case the assumed operator is ``non equal''
(\texttt{!=}).  Therefore, all these queries are equivalent to the one
written above
$$
\texttt{instrument ACS} \qquad
\texttt{=ACS} \qquad
\texttt{ACS}
$$
while \texttt{-ACS} is interpreted as \texttt{instrument\!ACS}.

\subsection{Automatic keyword identification}
\label{sec:autom-keyw-ident}

As discussed above, the entity identification is a key feature of the
IEAD system and a critical step toward a natural language query.  This
task is performed by attaching to each word (or group of words) a
keyword and operator, when these have not been explicitly assigned.

The automatic identification of the keyword uses a simple but
effective scheme.  The value is compared with a set of known values or
formats, and the first matching keyword is used.  Keywords are sorted
in a way that most used ones, or the most restrictive ones (i.e.\ the
ones that give a match only in rather specific cases) are at the
beginning, so that in the large majority of cases the automatic
identification is successful (see Table~\ref{tab:1} for a list of
accepted keywords).  Note, in particular, that the target name is the
last tried keyword: target names can have many different formats, and
additionally it is very time-consuming to verify if a phrase can be
taken to be a target name.  As mentioned in Sect.~\ref{sec:user-aids},
the user can verify in real time the operated identification, and if
necessary, in the rare cases where the identification is inappropriate
or where the keyword cannot be automatically recognized (see last
column of Table~\ref{tab:1}), specify the desired keyword.

\subsection{Formats for values}
\label{sec:formats-values}

A second key feature of the IEAD system is the treatment of values and
their identification.  This part of the parsing process influences the
automatic keyword identification and therefore has been designed with
particular care.  The currently possible kind of values include:
\begin{description}
\item[Word.] A single word can be used to specify, for example, an
  instrument, a filter, or a dataset ID.  Typically, these values are
  can be assigned an automatic keyword for quantities that can only
  assume a value among a limited list of values (such as the
  instruments) or that have a specific format (such as the dataset
  ID).
\item[Phrase.] A few values are composed by several words (cf.\ for
  example the \texttt{dataset} keyword in the examples of
  Sect.~\ref{sec:one-line-queries}).  These can be just placed one
  after the other, or can be enclosed within single or double quotes
  to avoid ambiguities.
\item[Number.] Integer or floating point values.  Floating point
  numbers can use the scientific notation
  $\mathit{d.ddd}\mathtt{e}{\pm}\mathit{dd}$, i.e.\ with the letter
  \texttt{e} as separator between the mantissa and the exponent.
\item[Number with unit.] A number \textit{immediately\/} followed by a
  unit \textit{without spaces\/} is used for many terms, such as
  exposure time or pixel scale.  Different units are allowed for the
  same quantity (for example seconds, minutes, hours for the exposure
  time; see Table~\ref{tab:2}).
\item[Angle.] Coordinates are entered in the format
  $\pm$\textit{ddd}\texttt{:}\textit{mm\/}\texttt{:}\textit{ss\/}
  (\textit{hh}\texttt{:}\textit{mm\/}\texttt{:}\textit{ss\/} for Right
  Ascension) or with optional decimals for the (arc)seconds; when a
  keyword is specified, one can alternatively use fractional degrees
  or minutes.  Without keywords, coordinates are always taken to be
  equatorial, and in particular coordinates without sign are
  interpreted as Right Ascention, and coordinates with sign are
  interpreted as Declination.
\item[Date.] Dates must be always entered in the format
  \textit{yyyy\/}\texttt{-}\textit{mm\/}\texttt{-}\textit{dd\/}.
\item[Range.] A range can be entered using the format
  \textit{min\/}\texttt{..}\textit{max\/}, where either \textit{min\/}
  or \textit{max\/} can be omitted (but not both).  Ranges are
  accepted for all cases where a numerical value is valid, including
  angles, dates, numbers, and numbers with unit (in this case, a unit
  entered only for one of the two extremes is applied to the other
  extreme too).  A range can only be used with the equality
  (\texttt{=}) or inequality (\texttt{!=}) operator, and is then
  converted internally into expressions such as
  \textit{keyword\/}\lb{>=}\textit{min\/} \texttt{and}
  \textit{keyword\/}\lb{<=}\textit{max\/} for the equality and
  \textit{keyword\/}\lb{<}\textit{min\/} \texttt{or}
  \textit{keyword\/}\lb{>}\textit{max\/} for the inequality operators.
\end{description}

\subsection{Special cases}
\label{sec:special-cases}

A few terms are interpreted in a special way to accommodate particular
cases.  Many of these are especially important because the interface
performs for them a \textit{query expansion\/}, i.e.\ it extents the
meaning of the human-entered values to adapt them to the database.
\begin{description}
\item[Target name.] All terms that cannot be automatically associated
  to any keyword are taken to be target names.  These are resolved in
  real time through a call to the Sesame name resolver, which by
  default queries in sequence the SIMBAD, NED, and VizieR
  \citep{2000A&AS..143...23O} databases.  The result of the Sesame
  check is immediately shown to the user as soon as it is available,
  typically within a couple of seconds.  The results are cached, so
  the same target is never queried again to Sesame within the same
  session.
\item[Coordinate pair.] Two \textit{close\/} coordinates terms (Right
  Ascension and Declination, or Galactic latitude and longitude, or
  Ecliptic latitude and longitude) are interpreted as a coordinate
  pair.  This is important when parsing positional constraints, since
  these are almost always taken with an implicit or explicit
  ``fuzziness'' (see below ``Search radius'').
\item[Search radius.] An isolated angle quantity (i.e., a number
  followed by an angular unit, such as \texttt{2arcsec}) is taken to
  be an indication for an angular resolution of the observation.
  However, when the same quantity appears close to a coordinate term
  (such as \texttt{12:32:45} with is parsed as a Right Ascension
  coordinate), to a coordinate pair (see above), or to a target name,
  it is taken to be a search radius for around the coordinate, the
  coordinate pair, or the target.  When not specified, all coordinates
  with (implicit or explicit) equality or inequality operator are
  taken to have a search radius of $10 \mbox{ arcmin}$.  Note also
  that a search radius works differently depending if it is applied to
  a point in the celestial sphere (a coordinate pair, see below) or to
  a single coordinate (in this case the search ``radius'' does not
  identifies a disk but rather a stripe in the sky).
\item[PI names.] Since different people have different habits for
  writing names, the system processes PI names so that the first name
  can be entered before or after the last name (more complicated cases
  where a middle name is present, or where the last name is composed
  by more words are also contemplated).  Internally the entered PI
  name is mapped into the format used by the database.  Additionally,
  for PIs the dot is equivalent to the \texttt{*} wildcard (see below
  Sect.~\ref{sec:wildcards-anchors}).
\item[Spectral range.] This is a special kind of number with unit,
  where the value entered is interpreted by requiring that the
  specific wavelength is \textit{within\/} the filter sensitivity of
  the dataset (or the specified wavelength range has a non-vanishing
  intersection with the filter sensitivity).  Additionally, the
  spectral range can also be entered using standard Johnson-Cousins
  filter names, which are approximately translated into the
  corresponding pivot wavelengths.  This feature, together with the
  capability of the system to recognize simple phrases for data types
  (such as \texttt{imaging} or \texttt{2d spectroscopy}) makes it
  possible to use the interface without a specific knowledge on the
  HST instrument capabilities.
\end{description}

\subsection{Wildcards and anchors}
\label{sec:wildcards-anchors}

IEAD also accepts two wildcards for text (non-numerical) values: the
asterisk (\texttt{*}), matching any text, and the question mark
(\texttt{?}), matching a single character.  These are the same
wildcards used in globbing by UNIX shells, and therefore should be
familiar to many users.

Additionally, the system also accept the caret (\verb|^|) to match the
beginning of a value, and the dollar sign (\verb|$|) to match the end
of a value.  Therefore, to force a keyword to have exactly a given
value one should use both anchors, as in \verb|title=^star$|.  Note
that the simple \verb|title=star| would match also observations with
titles such as ``A peculiar star in a nebula''.

Both wildcards and anchors can be escaped using the backslash, as in
\verb|\*| or \verb|\^|.

\subsection{Multiple constraints}
\label{sec:multiple-contraints}

Multiple constraints can be simply written one after the other.
Again, the specific framework of our query language, astronomical
databases, makes it simple to define rules that correspond to the
ones of a natural language:
\begin{itemize}
\item All terms with (implicit or explicit) equality operator that
  share the same (implicit or explicit) keyword are combined with an
  \texttt{or} boolean operator;
\item Other terms are combined with an \textrm{and} boolean operator.
\end{itemize}
Note that in different contexts this particular problem, i.e.\ the
so-called conjunction and disjunction ambiguity, is of difficult
solution because it requires a knowledge on the relationships among
the various entities, while in the astronomical context the solution
is obvious: all entities (for example, instruments, cameras, filters)
are mutually exclusive, in the sense that an observation can only use
one of them at a time.

These rules make sure that the simple query
$$
\texttt{m42 acs wfc3}
$$
is interpreted as a search for observations around M42 carried out 
\textit{either\/} with the ACS \textit{or\/} with the WFC3
instruments, while
$$
\texttt{m42 -acs -wfc3}
$$
is interpreted as a search for observations around M42 carried out
\textit{neither\/} with the ACS \textit{nor\/} with the WFC3
instruments.

\subsection{Full boolean queries}
\label{sec:full-boolean-queries}

The combination of the automatic identification and of the simple
syntax for combined constraints nicely solves the large majority of
queries.  However, in specific cases one might desire or need to
perform more specific queries involving combination of parameters.

In these situations it is possible to include the boolean operators
\texttt{and}, \texttt{or}, and \texttt{not} in the query and use them
as one would normally do in any programming language.  For example,
the string
$$
 \texttt{(acs and grism) or (stis and prism)}
$$
might represent a sensible query.  It is possible to mix multiple
contraints without boolean operators with queries involving boolean
operators: for example, the query 
$$
 \texttt{acs grism or stis prism}
$$
would be interpreted exactly as the example above (note that the
implicit boolean operators inserted between multiple constraints have
a higher priority than explicit boolean operators).

\section{Implementation}
\label{sec:implementation}

As mentioned earlier, IEAD is entirely written in JavaScript, so that
the code runs directly on the user's browser and can perform truly
interactive actions such as the automatic display of the parsing
result and the context-sensitive autocompletion.  Indeed, the choice
of the programming language for the final implementation has been
mainly driven by the possibility to display interactive messages to
the user (the original prototype of the interface was written in
Python).

The code is object oriented and is built over the concept of a term,
i.e.\ a combination of keyword-operator-value.  It defines a different
kind of object for each different type of term: number (integer or
float), angle, date, word, phrase, unit, flag, wavelength, pi (cf.\
Table~\ref{tab:1}).  All term classes are organized hierarchically
with a common ancestor.

Each object has a constructor, which defines the keyword (and
associated aliases) of the term, the associated field in the database,
and type-dependent options (for example, for the angle the allowed
range, the obligatoriness of the sign, and a flag to indicate if the
field refers to right ascension or not).  Additionally, all term
classes define a number of common members to deal with several common
actions, such as the verification of the input, the parsing, the error
handling, the autocompletion.  Finally, a member function takes care
of the translation of the term into SQL code or into a human readable
string.

The program also defines meta-terms, i.e.\ classes to modify the
behaviour of other terms.  For example there is a meta-term that
modifies other terms to make the keyword and/or the operator of a term
compulsory; another meta-term makes other terms optional (in the sense
that if not present, than a default value is used; cf.\ the use of the
search box, explained above).  However, the most important meta-term
is the one to generate ranges that use a pair of dots as separator.

Finally, all terms are combined together into a parser for an
expression that can involve boolean operators and parentheses: again
this process is realized within a special class with a structure
similar to the ones of a term.

In summary, the code uses the following simple scheme:
\begin{enumerate}
\item The query is initially handled by a simple lexer that splits the
  string into tokens, i.e.\ words that have an individual meaning in
  the language used by IEAD.
\item The tokens are passed to the expression parser, which analyses
  them in order.
\item The parser tries to parse the various terms by trying, in
  sequence, all term types that define an expression.  The first
  matching term is used for the rest of the parsing.  The last
  possible term tested is the target one, which by default accepts all
  tokens (unless a different keyword is specified).  When the target
  token is used, a query to the Sesame database is also started in
  parallel.
\item The prededing point is repeated until all tokens are consumed.  
\item The code uses the parser to translate the query into a human
  readable string that is shown to the user.
\item When the query is finally executed, the parser is also called
  again to generate an SQL code.  This code is used directly in a
  special field in an HTML form, and is then passed to the server which
  uses it to directly interrogate the database with very little
  manipulation.
\end{enumerate}

\subsection{Internal database}
\label{sec:internal-database}

The IEAD code uses a simple internal JSON database to save the values
for the various terms that have a fixed set of permitted values, such
as the instrument names or the filter names.  Periodically this
database is updated to reflect the status of the full HST archive:
this process is particularly important for values such as the PI
names, which are likely to change often (basically each time
observations from a new PI are carried out).

This simple process is handled by a straightforward Python script.
The only interesting point to note here is that the script performs
the same analysis for the \texttt{description} term.  Since (almost)
each proposal has a different description, often containing several
words, it would be unpractical and probably not very useful to use
these values as they are for the \texttt{description} term.  Rather,
the Python script analyses the various descriptions, and extracts from
them the most common phrases.  These are automatically
recognized and also used for autocompletion in the interface.  Of
course, one is still free to query for a particular description using
a full qualified term, as in \texttt{description}\=\textit{phrase\/}.

\section{Future prospects}
\label{sec:future-prospects}

The research presented here is the first step toward more advanced,
efficient, and intuitive query interfaces for astronomical databases.
However, this should not be considered the definitive, optimal solution.
Instead, given the role that astronomical databases will play in the
future astronomical research, it is critical to develop even more
advanced interfaces to fulfill the needs of the users and to stimulate
different uses of the astronomical databases.  The interface discussed
in this paper could be improved in various ways, and it is useful to
briefly consider future research directions here.

The interface could provide instant previews of the entered query,
without forcing the user to press the ``Search'' button each time.
This would allow one to explore in real time the archive and as such
would represent a major step forward.  Unfortunately, at the present
the structure and the query performance of most archives does not
allow such a search to be performed in real time, and major
enhancements in the database software and servers would be needed for
this task.

Improvements could also be made in the auto-completion scheme.  So
far, the autocompletion system presents a truncated (or, optionally,
the full) list of possible completions for a query, but by no means it
tries to perform a ``smart'' job.  First, the list produced is
presented in alphabetical order, which is not the most useful one (it
would be much more sensible to present the list of completions by
sorting them by ``popularity,'' i.e.\ frequency of use).  Second, the
autocompletion is only grammatically context-sensitive, not
semantically: the proposed completions are likely to contain
possibilities that would be considered obviously wrong by an
experienced user (for example, the interface would include the
\texttt{WFPC2} instrument as a possible completion of ``\texttt{WF}''
even if the user already selected the grism \texttt{G280}, which is
not available for this instrument).  This improvement, however,
appears to be rather complicated to develop.

As mentioned already, the new interface implements many of the tasks
needed in a \textit{natural language query\/}, such as entity
identification, query expansion, and conjunction-disjunction
disambiguation.  The ad-hoc implementation is computationally
efficient, but has also various limitations: the set of keywords must
be provided to the system together with a way to discover from the
database the set of allowed values; and the query language is kept at
a very basic level.  There are various possible solutions for these
issues.  The need for a specific configuration of the system (keywords
and predefined values) could be removed through the use of VO
registries \citep{VOReg}: this would make it possible to port the
interface to all VO-compliant archives, possibly even without the
explicit collaboration from the maintainers.  Regarding possible
extensions of the query language, it would be very interesting to
investigate modern techniques used in natural language research,
involving statistical inference and machine learning.  However, one
should also be aware of the possible risks of a full natural language
query: on the one hand, the lack of a well defined grammar, and
therefore of the set of possible queries, might keep the users away
from queries that are considered too complex; on the other hand, many
users would expect the system to be ``intelligent'' and might request
queries that are outside of its capabilities.

\section{Conclusions}
\label{sec:conclusions}

In this paper I presented IEAD, a new one-line query interface for
astronomical science databases.  The major advantages of this
interface over standard ones are
\begin{itemize}
\item Queries are performed on a single line.  This makes query pages
  very clean, avoid the clumsiness often present in astronomical
  archive interfaces, and let the user concentrate on the query
  (instead of on the search of right field of each constrain).
\item The interface uses a simple syntax, designed to minimize the
  quantity of text the user has to enter and to be close to a natural
  language.
\item Queries involving complex combination of parameters, boolean
  operators, and parentheses are possible.
\item The interface provides immediate feedback on the parsing of the
  entered string and on the resolution of astronomical object names.
  Additionally, it has context sensitive autocompletion.
\item The code is easily integrable in any SQL-compliant astronomical
  database, is extensible and usable for different telescopes or
  observatories.
\end{itemize}

\acknowledgments

I am grateful to F.~Stoehr and J.~Haase for helpful and constructive
interactions, and to R.~Fosbury and J.R.~Walsh for encouraging and
supporting this project.

\bibliographystyle{apj} 
\bibliography{ms}

\begin{thebibliography}{11}
\expandafter\ifx\csname natexlab\endcsname\relax\def\natexlab#1{#1}\fi

\bibitem[{{Androutsopoulos} {et~al.}(1995){Androutsopoulos}, {Ritchie}, \&
  {Peter}}]{ART95}
{Androutsopoulos}, I., {Ritchie}, G.~D., \& {Peter}, T. 1995, Natural Language
  Engineering, 1, 29

\bibitem[{{Benson} {et~al.}(2009){Benson}, {Plante}, {Auden}, {Graham},
  {Greene}, {Hill}, {Linde}, {Morris}, {O’Mullane}, {Rixon}, {Stébé}, \&
  {Andrews}}]{VOReg}
{Benson}, K., {Plante}, R., {Auden}, E., {Graham}, M., {Greene}, G., {Hill},
  M., {Linde}, T., {Morris}, D., {O’Mullane}, W., {Rixon}, G., {Stébé}, A.,
  \& {Andrews}, K. 2009, {IVOA} Registry Interface, Tech. rep., International
  Virtual Observatory Alliance,
  \url{http://www.ivoa.net/Documents/RegistryInterface/}

\bibitem[{{Delmotte} {et~al.}(2006){Delmotte}, {Dolensky}, {Micol}, {Padovani},
  {Rino}, {Rosati}, {Wicenec}, {Retzlaff}, {Rit{\'e}}, {Slijkhuis}, {Vandame},
  \& {Vuong}}]{2006Msngr.125...41D}
{Delmotte}, N., {Dolensky}, M., {Micol}, A., {Padovani}, P., {Rino}, B.,
  {Rosati}, P., {Wicenec}, A., {Retzlaff}, J., {Rit{\'e}}, C., {Slijkhuis}, R.,
  {Vandame}, B., \& {Vuong}, M.~H. 2006, The Messenger, 125, 41

\bibitem[{{Helou} {et~al.}(1991){Helou}, {Madore}, {Schmitz}, {Bicay}, {Wu}, \&
  {Bennett}}]{1991ASSL..171...89H}
{Helou}, G., {Madore}, B.~F., {Schmitz}, M., {Bicay}, M.~D., {Wu}, X., \&
  {Bennett}, J. 1991, in Astrophysics and Space Science Library, Vol. 171,
  Databases and On-line Data in Astronomy, ed. {M.~A.~Albrecht \& D.~Egret},
  89--106

\bibitem[{{Ochsenbein} {et~al.}(2000){Ochsenbein}, {Bauer}, \&
  {Marcout}}]{2000A&AS..143...23O}
{Ochsenbein}, F., {Bauer}, P., \& {Marcout}, J. 2000, \aaps, 143, 23

\bibitem[{{Ortiz} {et~al.}(2008){Ortiz}, {Lusted}, {Dowler}, {Szalay},
  {Shirasaki}, {NietoSantisteban}, {Ohishi}, {O'Mullane}, Pedro, {the VOQL-TEG
  team}, \& {the VOQL Working Group}.}]{ADQL}
{Ortiz}, I., {Lusted}, J., {Dowler}, P., {Szalay}, A., {Shirasaki}, Y.,
  {NietoSantisteban}, M.~A., {Ohishi}, M., {O'Mullane}, W., Pedro, O., {the
  VOQL-TEG team}, \& {the VOQL Working Group}. 2008, {IVOA} Astronomical Data
  Query Language, Tech. rep., International Virtual Observatory Alliance,
  \url{http://www.ivoa.net/Documents/latest/ADQL.html}

\bibitem[{{Popescu} {et~al.}(2003){Popescu}, {Etzioni}, \& {Kautz}}]{PEK03}
{Popescu}, A.-M., {Etzioni}, O., \& {Kautz}, H. 2003, in Proceedings of the 8th
  international conference on intelligent user interfaces, {IUI} '03 (New York,
  NY, USA: ACM), 149--157

\bibitem[{{Skrutskie} {et~al.}(2006){Skrutskie}, {Cutri}, {Stiening},
  {Weinberg}, {Schneider}, {Carpenter}, {Beichman}, {Capps}, {Chester},
  {Elias}, {Huchra}, {Liebert}, {Lonsdale}, {Monet}, {Price}, {Seitzer},
  {Jarrett}, {Kirkpatrick}, {Gizis}, {Howard}, {Evans}, {Fowler}, {Fullmer},
  {Hurt}, {Light}, {Kopan}, {Marsh}, {McCallon}, {Tam}, {Van Dyk}, \&
  {Wheelock}}]{2006AJ....131.1163S}
{Skrutskie}, M.~F., {Cutri}, R.~M., {Stiening}, R., {Weinberg}, M.~D.,
  {Schneider}, S., {Carpenter}, J.~M., {Beichman}, C., {Capps}, R., {Chester},
  T., {Elias}, J., {Huchra}, J., {Liebert}, J., {Lonsdale}, C., {Monet}, D.~G.,
  {Price}, S., {Seitzer}, P., {Jarrett}, T., {Kirkpatrick}, J.~D., {Gizis},
  J.~E., {Howard}, E., {Evans}, T., {Fowler}, J., {Fullmer}, L., {Hurt}, R.,
  {Light}, R., {Kopan}, E.~L., {Marsh}, K.~A., {McCallon}, H.~L., {Tam}, R.,
  {Van Dyk}, S., \& {Wheelock}, S. 2006, \aj, 131, 1163

\bibitem[{{Walsh} \& {Hook}(2006)}]{2006STECF..40....6W}
{Walsh}, J. \& {Hook}, R. 2006, Space Telescope European Coordinating Facility
  Newsletter, 40, 6

\bibitem[{{Wenger} {et~al.}(2000){Wenger}, {Ochsenbein}, {Egret}, {Dubois},
  {Bonnarel}, {Borde}, {Genova}, {Jasniewicz}, {Lalo{\"e}}, {Lesteven}, \&
  {Monier}}]{2000A&AS..143....9W}
{Wenger}, M., {Ochsenbein}, F., {Egret}, D., {Dubois}, P., {Bonnarel}, F.,
  {Borde}, S., {Genova}, F., {Jasniewicz}, G., {Lalo{\"e}}, S., {Lesteven}, S.,
  \& {Monier}, R. 2000, \aaps, 143, 9

\bibitem[{{York} {et~al.}(2000){York}, {Adelman}, {Anderson}, {Anderson},
  {Annis}, {Bahcall}, {Bakken}, {Barkhouser}, {Bastian}, {Berman}, {Boroski},
  {Bracker}, {Briegel}, {Briggs}, {Brinkmann}, {Brunner}, {Burles}, {Carey},
  {Carr}, {Castander}, {Chen}, {Colestock}, {Connolly}, {Crocker}, {Csabai},
  {Czarapata}, {Davis}, {Doi}, {Dombeck}, {Eisenstein}, {Ellman}, {Elms},
  {Evans}, {Fan}, {Federwitz}, {Fiscelli}, {Friedman}, {Frieman}, {Fukugita},
  {Gillespie}, {Gunn}, {Gurbani}, {de Haas}, {Haldeman}, {Harris}, {Hayes},
  {Heckman}, {Hennessy}, {Hindsley}, {Holm}, {Holmgren}, {Huang}, {Hull},
  {Husby}, {Ichikawa}, {Ichikawa}, {Ivezi{\'c}}, {Kent}, {Kim}, {Kinney},
  {Klaene}, {Kleinman}, {Kleinman}, {Knapp}, {Korienek}, {Kron}, {Kunszt},
  {Lamb}, {Lee}, {Leger}, {Limmongkol}, {Lindenmeyer}, {Long}, {Loomis},
  {Loveday}, {Lucinio}, {Lupton}, {MacKinnon}, {Mannery}, {Mantsch}, {Margon},
  {McGehee}, {McKay}, {Meiksin}, {Merelli}, {Monet}, {Munn}, {Narayanan},
  {Nash}, {Neilsen}, {Neswold}, {Newberg}, {Nichol}, {Nicinski}, {Nonino},
  {Okada}, {Okamura}, {Ostriker}, {Owen}, {Pauls}, {Peoples}, {Peterson},
  {Petravick}, {Pier}, {Pope}, {Pordes}, {Prosapio}, {Rechenmacher}, {Quinn},
  {Richards}, {Richmond}, {Rivetta}, {Rockosi}, {Ruthmansdorfer}, {Sandford},
  {Schlegel}, {Schneider}, {Sekiguchi}, {Sergey}, {Shimasaku}, {Siegmund},
  {Smee}, {Smith}, {Snedden}, {Stone}, {Stoughton}, {Strauss}, {Stubbs},
  {SubbaRao}, {Szalay}, {Szapudi}, {Szokoly}, {Thakar}, {Tremonti}, {Tucker},
  {Uomoto}, {Vanden Berk}, {Vogeley}, {Waddell}, {Wang}, {Watanabe},
  {Weinberg}, {Yanny}, {Yasuda}, \& {SDSS Collaboration}}]{2000AJ....120.1579Y}
{York}, D.~G., {Adelman}, J., {Anderson}, Jr., J.~E., {Anderson}, S.~F.,
  {Annis}, J., {Bahcall}, N.~A., {Bakken}, J.~A., {Barkhouser}, R., {Bastian},
  S., {Berman}, E., {Boroski}, W.~N., {Bracker}, S., {Briegel}, C., {Briggs},
  J.~W., {Brinkmann}, J., {Brunner}, R., {Burles}, S., {Carey}, L., {Carr},
  M.~A., {Castander}, F.~J., {Chen}, B., {Colestock}, P.~L., {Connolly}, A.~J.,
  {Crocker}, J.~H., {Csabai}, I., {Czarapata}, P.~C., {Davis}, J.~E., {Doi},
  M., {Dombeck}, T., {Eisenstein}, D., {Ellman}, N., {Elms}, B.~R., {Evans},
  M.~L., {Fan}, X., {Federwitz}, G.~R., {Fiscelli}, L., {Friedman}, S.,
  {Frieman}, J.~A., {Fukugita}, M., {Gillespie}, B., {Gunn}, J.~E., {Gurbani},
  V.~K., {de Haas}, E., {Haldeman}, M., {Harris}, F.~H., {Hayes}, J.,
  {Heckman}, T.~M., {Hennessy}, G.~S., {Hindsley}, R.~B., {Holm}, S.,
  {Holmgren}, D.~J., {Huang}, C.-h., {Hull}, C., {Husby}, D., {Ichikawa},
  S.-I., {Ichikawa}, T., {Ivezi{\'c}}, {\v Z}., {Kent}, S., {Kim}, R.~S.~J.,
  {Kinney}, E., {Klaene}, M., {Kleinman}, A.~N., {Kleinman}, S., {Knapp},
  G.~R., {Korienek}, J., {Kron}, R.~G., {Kunszt}, P.~Z., {Lamb}, D.~Q., {Lee},
  B., {Leger}, R.~F., {Limmongkol}, S., {Lindenmeyer}, C., {Long}, D.~C.,
  {Loomis}, C., {Loveday}, J., {Lucinio}, R., {Lupton}, R.~H., {MacKinnon}, B.,
  {Mannery}, E.~J., {Mantsch}, P.~M., {Margon}, B., {McGehee}, P., {McKay},
  T.~A., {Meiksin}, A., {Merelli}, A., {Monet}, D.~G., {Munn}, J.~A.,
  {Narayanan}, V.~K., {Nash}, T., {Neilsen}, E., {Neswold}, R., {Newberg},
  H.~J., {Nichol}, R.~C., {Nicinski}, T., {Nonino}, M., {Okada}, N., {Okamura},
  S., {Ostriker}, J.~P., {Owen}, R., {Pauls}, A.~G., {Peoples}, J., {Peterson},
  R.~L., {Petravick}, D., {Pier}, J.~R., {Pope}, A., {Pordes}, R., {Prosapio},
  A., {Rechenmacher}, R., {Quinn}, T.~R., {Richards}, G.~T., {Richmond}, M.~W.,
  {Rivetta}, C.~H., {Rockosi}, C.~M., {Ruthmansdorfer}, K., {Sandford}, D.,
  {Schlegel}, D.~J., {Schneider}, D.~P., {Sekiguchi}, M., {Sergey}, G.,
  {Shimasaku}, K., {Siegmund}, W.~A., {Smee}, S., {Smith}, J.~A., {Snedden},
  S., {Stone}, R., {Stoughton}, C., {Strauss}, M.~A., {Stubbs}, C., {SubbaRao},
  M., {Szalay}, A.~S., {Szapudi}, I., {Szokoly}, G.~P., {Thakar}, A.~R.,
  {Tremonti}, C., {Tucker}, D.~L., {Uomoto}, A., {Vanden Berk}, D., {Vogeley},
  M.~S., {Waddell}, P., {Wang}, S.-i., {Watanabe}, M., {Weinberg}, D.~H.,
  {Yanny}, B., {Yasuda}, N., \& {SDSS Collaboration}. 2000, \aj, 120, 1579

\end{thebibliography}

\clearpage

\clearpage

\begin{table*}
\begin{center}
\caption{List of currently accepted keywords and aliases.\label{tab:1}}
\begin{tabular}{llccc}
  \tableline\tableline
  Keyword & Aliases & Type & Units & Automatic \\
  \tableline
  \texttt{inst} & \texttt{instr}, \texttt{instrument} & word & --- &
  yes \\
  \texttt{camera} & \texttt{detector} & word & --- & yes \\
  \texttt{filter} & \texttt{opt\_elem}, \texttt{optical\_element} &
  word & --- & yes \\
  \texttt{opt\_elem\_type} & \texttt{optical\_element\_type} & word & --- &
  yes \\
  \texttt{data\_type} & --- & phrase & --- & yes \\
  \texttt{date} & \texttt{obsdate}, \texttt{observation\_date} & date
  & --- & yes \\
  \texttt{released} & \texttt{release\_date}, \texttt{observation\_date} & date
  & --- & no \\
  \texttt{ra} & --- & angle\tablenotemark{a} & ---  & yes \\
  \texttt{dec} & \texttt{decl}, \texttt{declination} &
  angle\tablenotemark{a} & ---  & yes \\
  \texttt{glon} & \texttt{galactic\_longitude}, \texttt{gal\_lon} &
  angle\tablenotemark{b} & ---  & yes \\ 
  \texttt{glat} & \texttt{galactic\_latitude}, \texttt{gal\_lat} &
  angle\tablenotemark{b} & ---  & yes \\
  \texttt{elon} & \texttt{ecliptic\_longitude}, \texttt{ecl\_lon} & angle\tablenotemark{c} & ---  & yes \\
  \texttt{elat} & \texttt{ecliptic\_latitude}, \texttt{ecl\_lat} & angle\tablenotemark{c} & ---  & yes \\
  \texttt{box} & \texttt{radius}, \texttt{r}, \texttt{within},
  \texttt{search\_box} & unit & angle & yes\tablenotemark{d} \\
  \texttt{hst\_target} & \texttt{hst\_target\_name},
  \texttt{hst\_name} & phrase & --- & no \\
  \texttt{description} & \texttt{descr}, \texttt{target\_description},
  & phrase & --- & yes\tablenotemark{e} \\
  & \texttt{targetdescription} \\
  & \texttt{descript}, \texttt{targetdescription} \\
  \texttt{exptime} & \texttt{exposure}, \texttt{exposure\_time} & unit
  & time & yes \\
  \texttt{prop} & \texttt{proposal}, \texttt{proposalid}, & integer &
  --- & yes \\
  & \texttt{proposal\_id}, \texttt{prop\_id} & & & \\
  \texttt{pi} & \texttt{pi\_name}, \texttt{piname},
  \texttt{principal\_investigator} & pi & --- & yes \\
  \texttt{dataset} & \texttt{dataset\_name}, \texttt{data\_set\_name}
  & word & --- & yes \\
  \texttt{title} & \texttt{proposal\_title}, \texttt{prop\_title} &
  phrase & --- & no \\
  \texttt{resolution} & \texttt{spatial\_resolution} & unit & arcsec &
  yes\tablenotemark{d} \\
  \texttt{scale} & \texttt{pixel\_scale}, \texttt{pixel} & unit &
  arcsec & no \\
  \texttt{slew} & \texttt{moving}, \texttt{moving\_object},
  \texttt{moving\_target} & flag & --- & no \\
  \texttt{wavelength} & \texttt{wave}, \texttt{lambda} & wavelength &
  wavelength\tablenotemark{f} & yes \\
  \texttt{bandwidth} & --- & unit & wavelength & no \\
  \texttt{spec\_res} & \texttt{spectral\_resolution} & unit &
  wavelength & no \\
  \texttt{res\_power} & \texttt{resolving\_power}, \texttt{respower} &
  float & --- & yes \\
  \texttt{time\_start} & \texttt{start} & date & --- & no \\
  \texttt{time\_end} & \texttt{end} & date & --- & no \\
  \texttt{members} & \texttt{no\_members} & integer & --- & no \\
  \texttt{mode} & \texttt{photon\_mode} & phrase & --- & no \\
  \texttt{extension} & \texttt{science\_extension} & phrase & --- &
  yes \\
  \tableline
\end{tabular}
\tablenotetext{a,b,c}{Coordinate pairs.}  \tablenotetext{d}{The
  \texttt{box} term is valid only near a coordinate, and in that case
  takes priority over \texttt{resolution}.  Therefore,
  \texttt{resolution} is automatic only when there is no nearby
  coordinate.}
\tablenotetext{e}{The \texttt{description} is automatically recognized
  for the most common description phrases, as deduced from a periodic
  analysis of the proposal database; typical examples include
  ``\texttt{star forming region}'' or ``\texttt{cluster of
    galaxies}''.}
\tablenotetext{f}{In addition to wavelength units, the following
  astronomical bands are also recognized: \texttt{ultraviolet},
  \texttt{optical}, \texttt{infrared}, \texttt{u}, \texttt{b},
  \texttt{g}, \texttt{v}, \texttt{r}, \texttt{i}, \texttt{z},
  \texttt{j}, \texttt{h}, and \texttt{k}.}
\end{center}
\end{table*}

\clearpage

\begin{table*}
\begin{center}
\caption{List of unit types used for different keywords (cf.\
  Table~\ref{tab:1}).\label{tab:2}}
\begin{tabular}{llll}
  \tableline\tableline
  Unit type & Default & Units & Factor \\
  \tableline
  Angle & \texttt{arcsec} & \texttt{arcsec, arcsecond, arcseconds} &
  1/3600 \\
  & & \texttt{arcmin, arcminute, arcminutes} & 1/60 \\
  & & \texttt{deg, degree, degrees} & 1 \\
  \tableline
  Time & \texttt{s} & \texttt{s, sec, second, seconds} & 1 \\
  & & \texttt{m, min, minute, minutes} & 60 \\
  & & \texttt{h, hour, hours} & 3600 \\
  \tableline
  Wavelength & \texttt{nm} & \texttt{nm, nanometer, nanometers} & 1 \\
  & & \texttt{a, ang, angstrom, angstroms} & 1/10 \\
  & & \texttt{um, micron, microns} & 1000 \\
  \tableline
\end{tabular}
\end{center}
\end{table*}

\end{document}